\documentclass[fleqn,usenatbib]{mnras}
\usepackage{amsmath}	
\usepackage{amssymb}	
\providecommand{\abs}[1]{\lvert#1\rvert}
\usepackage{newtxtext,newtxmath}

\usepackage[T1]{fontenc}
\usepackage{ae,aecompl}
\usepackage{hyperref}
\usepackage[]{lineno}
\usepackage{threeparttable}
\usepackage{cancel}
\modulolinenumbers[2]

\usepackage{float}
\usepackage{graphicx}	

\usepackage{xcolor}
\usepackage{tikz}

\definecolor{awesome}{rgb}{1.0, 0.13, 0.32}
\definecolor{lightcyan}{rgb}{0.88,1,1}
\definecolor{bluegray}{rgb}{0.4, 0.6, 0.8}

\definecolor{lime}{HTML}{A6CE39}
\DeclareRobustCommand{\orcidicon}{%
    \hspace{-2mm}
	\begin{tikzpicture}
	\draw[lime, fill=lime] (0,0) 
	circle [radius=0.16] 
	node[white] {{\fontfamily{qag}\selectfont \tiny ID}};
	\draw[white, fill=white] (-0.0625,0.095) 
	circle [radius=0.007];
	\end{tikzpicture}
	\hspace{-2mm}
}
\foreach \x in {A, ..., Z}{%
	\expandafter\xdef\csname orcid\x\endcsname{\noexpand\href{https://orcid.org/\csname orcidauthor\x\endcsname}{\noexpand\orcidicon}}
}

\DeclareRobustCommand{\VAN}[3]{#2}
\let\VANthebibliography\thebibliography
\def\thebibliography{\DeclareRobustCommand{\VAN}[3]{##3}\VANthebibliography}

\title[Capture and migration of Jupiter and Saturn in MMR]{Capture and migration of Jupiter and Saturn in mean motion resonance in a gaseous protoplanetary disc}

\author[Chametla et al.]{%
Ra\'ul O. Chametla,$^{1}$\thanks{E-mail: rortegaesfm@gmail.com (RO); gennaro@lanl.gov (GD); maurey@astro.unam.mx (MR); jsanchez@astro.unam.mx (JS)}\orcidA{}
Gennaro D'Angelo,$^{2}$\footnotemark[1]\orcidB{}
Mauricio Reyes-Ruiz$^{3}$\footnotemark[1]\orcidC{} and
\newauthor 
F. Javier S\'anchez-\-Salcedo$^{4}$\footnotemark[1]\orcidD{}
\\
$^{1}$Instituto de Ciencias F\'isicas, Universidad Nacional Autonoma de M\'exico, Av. Universidad s/n, 62210 Cuernavaca, Mor., Mexico\\
$^{2}$Theoretical Division, Los Alamos National Laboratory, Los Alamos, NM 87545, USA\\
$^{3}$Universidad Nacional Aut\'onoma de M\'exico, Instituto de Astronom\'{\i}a, Ensenada,
B.C, Mexico\\
$^{4}$Instituto de Astronom\'{\i}a, Universidad Nacional Aut\'onoma de M\'exico, Ciudad
Universitaria, Apt. Postal 70-264, C.P. 04510,\\
Mexico City, Mexico\\
}

\date{Accepted XXX. Received YYY; in original form ZZZ}

\pubyear{2020}

\begin{document}
\label{firstpage}
\pagerange{\pageref{firstpage}--\pageref{lastpage}}
\maketitle

\begin{abstract}
We study the dynamical evolution of Jupiter and Saturn embedded in a gaseous, 
solar-nebula-type disc by means of hydrodynamics simulations with the FARGO2D1D code. 
We study the evolution for different initial separations of the planets' orbits, 
$\Delta a_{SJ}$, to investigate whether they become captured in mean motion 
resonance (MMR) and the direction of the subsequent migration of the planet 
(inward or outward).
We also provide an assessment of the planet's orbital dynamics at different 
epochs of Saturn's growth.
We find that the evolution of initially compact orbital 
configurations is dependent on the value of $\Delta a_{SJ}$. 
This implies that an evolution as that proposed in the Grand Tack model depends 
on the precise initial orbits of Jupiter and Saturn and on the timescales 
for their formation. 
Capture in the 1:2 MMR and inward or (nearly) stalled migration are highly favoured.
Within its limits,
our work suggests that the reversed migration, associated with the 
resonance capture of Jupiter and Saturn, may be a low probability evolutionary 
scenario, so that other planetary systems with giant planets are not expected 
to have experienced a Grand Tack-like evolutionary path. 
   
\end{abstract}




\section{Introduction}
 \label{sec:intro}

Over the past two decades, our understanding of the formation of the solar system has
changed significantly. For example, the ``Nice model'' \citep[][]{Tsiganisetal2005} 
was introduced to account for several aspects of the current architecture of the 
outer solar system, such as the possible chaotic capture of  the Trojan populations 
of Jupiter and Neptune \citep[][]{Morbidellietal2005} and the so-called Late Heavy 
Bombardment of the terrestrial planets \citep[see][]{Gomesetal2005}.
The model also provides a semi-quantitative description of the formation of the 
Kuiper Belt \citep[see][]{Levisonetal2008} and of the secular dynamics of the outer 
solar system \citep[][]{Morbidellietal2009a}. 
It should be noted, however, that 
this model has undergone several revisions 
\citep[see][and references therein]{Morbidelli2019} since it was first proposed.
Another example is the so-called ``Grand Tack'' scenario \citep[][]{Walshetal2011},
which aims to explain some properties of the inner solar system.
In this 
model, Jupiter would have formed around the snow line location in the gaseous stage 
of the protoplanetary disc \citep[][]{CieslaCuzzi2006}, approximately $3.5$ AU from the Sun,
where an increase in the surface density of solids (due mostly to condensation of ice) 
is believed to have led to favourable conditions for the formation of a giant planet. 
According to this scenario, around the time that Jupiter terminated its rapid accretion 
of gas from the disc, Saturn was still growing, having reached about $30$ percent of its 
current mass and being located at $\approx 4.5$ AU from the Sun. 
While Saturn was completing the gas accretion phase, to reach its final mass 
($\approx 95\, M_{\oplus}$), it started to migrate inward at a rate faster than Jupiter,
eventually reaching the location of Jupiter's outer 3:2 mean motion resonance (MMR). 
At this point during their evolution, Jupiter had reached a distance of $\approx 1.5$~AU 
from the Sun and Saturn had acquired a mass of approximately $60\, M_{\oplus}$.
Thereafter, the orbits of the pair remained locked in the resonant configuration. 

In the Grand Tack scenario, capture into resonance led the migration of Jupiter and Saturn 
to reverse and proceed outward, while maintaining resonance locking. After a period of 
$\sim 2\times10^5$ yr, during which the planets drifted outward in the disc, they eventually 
reached the recently formed Uranus and Neptune, located at about $6$ and $8$~AU, respectively, 
and they too were captured in resonances and pushed outwards, away from the Sun. Outward 
migration eventually stopped as the gaseous disc dissipated and the solar system was left 
in a compact, multi-resonant configuration.
The final stage shaping the current orbital architecture of the solar system involved 
the interaction of the outer planets with the remnant disc of planetesimals, which led to 
a chaotic and rapid outward migration to their present day configuration \citep[][]{Tsiganisetal2005,Walshetal2011}. 
This scenario has been used to explain the orbital evolution and composition of the asteroid 
belt in conjunction with Jupiter's migration, a complex issue in the context of the 
classical Nice model. 

The key ingredient of the Grand Tack is the migration process driven by tidal interactions 
with the gaseous disc and the assumed migration reversal, which leads to a compact multi-resonant 
configuration of the giant planets in the early solar system.
Had Jupiter not interacted with another planet, or with a massive disc of planetesimals, 
once it had opened a gap in the gas disc it would have migrated towards the 
central star driven by the disc's tidal torques, in what is commonly referred 
to as classical type~II migration \citep[and references therein]{Lin1993}.
Although the migration regime of gap-opening planets in more realistic discs has been reconsidered
over the years 
\citep[e..g.,][]{Gennaro2008,Gennaro2010,Duffell2014,Duermann2015,Kanagawa2018}, 
it would still result in inward migration.
However, gap formation and ensuing dynamical evolution of a massive 
planet in the disc can be modified if there is another planet perturbing 
the disc \citep{Kley2012}. 
Such gravitational interaction can slow down, halt, or even reverse the migration of Jupiter 
\citep[][]{MassetSnellgrove2001,MorbidelliCrida2007,ZhangZhou2010,Gennaro2012}.
The resulting multi-resonant configurations depend on the initial conditions, the disc viscosity 
$\nu$, the disc aspect ratio $h = H/r$, and on the surface density profile of the disc $\Sigma(r)$. 
The Grand Tack was inspired by \citet{MassetSnellgrove2001}, who studied the migration of 
Jupiter and Saturn initially located at $5.2$ and $10.4$~AU, respectively, and embedded in 
a disc with $h=0.04$, uniform $\Sigma$, and uniform kinematic viscosity
corresponding to a value of the turbulence parameter $\alpha=10^{-3}$, 
where $\nu\propto \alpha h^{2}$ in the prescription of \citet{SakuraSunyaev1973}.

\citet{MorbidelliCrida2007} explored the effect of different disc parameters (such as $h$ and $\nu$) 
and found that the outward migration of Jupiter and Saturn occurs if $\nu\approx 10^{-6}$ 
and $h\approx 0.03$. With these parameters, Jupiter and Saturn can get trapped in the 3:2 MMR. 
However, there are several combinations of $\nu$ and $h$ that lead to a quasi-stationary 
evolution (i.e., very slow migration) once the planets are caught in resonance. 
This occurs when the ratio between the inner and the outer planet masses is less than $3$.

\citet{ZhangZhou2010} considered the effect of the surface density profile on the migration of Jupiter and Saturn. For a surface density $\Sigma(r) \propto r^{-\beta}$ 
(and fixed $\nu=10^{-6}$ and $h=0.04$), they found that when $0<\beta<1$, Jupiter and Saturn 
are trapped into 2:1 MMR because their convergent migration speed is too slow to break through 
the resonance. For $\beta>4/3$, Saturn and Jupiter get locked into 3:2 MMR. Interestingly, 
they argue that Saturn's gap is modified by the tidal perturbation of Jupiter, so that Saturn 
can migrate outwards or inwards. They found that outward migration occurs for $\beta\geq 4/3$, and inward migration for $\beta\leq 1$. 
In a subsequent work, \citet{Zhang-Zhou2010b} studied the orbital evolution 
of a planet pair where the outer planet is more massive than the inner one. 
They found that convergent migration occurs when the initial orbital separation 
between them is relatively small and $\beta<0.5$.

\citet{Gennaro2012} quantified the relative migration velocity, and the corresponding disc conditions 
($\Sigma$, $h$, $\nu$) that may lead to orbital trapping in the 3:2 MMR and outward migration. In addition, they also pointed out that ongoing accretion of gas onto the planets may deactivate the outward 
migration by changing the mass ratio, $M_J/M_S$, and/or by depleting the inner disc. 
They showed that typical conditions in evolved discs highly favour capture in the 2:1 resonance.

\citet{Pierens2014} also found that Jupiter and Saturn become trapped in the 2:1 
MMR, for gas densities $\lesssim 1800$~g~cm$^{-2}$ at $1$~AU, and that
they may migrate outward in low-viscosity discs, $\alpha \le 10^{-4}$, with very 
small aspect ratios, $h \sim 0.02$. The authors argued that such small pressure
scale-heights are possible at late times during disc evolution 
\citep[see also][]{Gennaro2012}, which would also imply that the planets evolved 
in resonance in a rather low-density nebula.

In the work carried out to date on the migration of Jupiter and Saturn trapped 
in resonance, the effects of the initial distance between the orbits of the two 
planets have not been analysed in the context of the Grand Tack,
although the initial separation may represent a key parameter for the onset 
of outward migration
since it directly influences the relative velocity of approach and hence the resulting MMR. In
this regard, it should be stressed that the orbits of the giant planets at the time of their 
formation are not well constrained, with proposed values differing by as much as 50\% 
\citep{Kley2000,LeePeale2002,Kleyetal2005}.
The initial separations of the giant planets in the solar system is also currently debated 
\citep{Nesvorny2018}.
In this paper, we investigate the effects of the initial (orbital) positions of Jupiter and Saturn 
on the process of resonant capture, during the gas-dominated stage of the protoplanetary disc. 
We consider different initial orbital separations between the planets, 
$\Delta a_{SJ} = r_S - r_J$ (where $r_J$ and $r_S$ are the initial semi-major axis of Jupiter 
and Saturn, respectively). We investigate how the direction and extent of Jupiter's migration 
depend on $\Delta a_{SJ}$. 

Among the models considered herein, those that most closely replicate
the assumptions invoked by the Grand Tack scenario result
in capture of Saturn in the 1:2 mean motion resonance with Jupiter.
Thereafter, the migration of the pair of planets is either stalled
or directed inwards.

The paper is organised as follows. In Section~\ref{sec:init_cond}, 
we describe the models. In Section~\ref{sec:results}, we report on 
the results of the simulations and discuss their implications. 
In Section~\ref{sec:conclusions}, we deliver the main conclusions.

\section{DESCRIPTION OF THE CODE AND INITIAL CONDITIONS}
\label{sec:init_cond}
We consider a thin, two-dimensional disc and adopt a polar coordinate system $(r,\phi)$, with 
the origin attached to the centre of mass of the system (star plus planets).
Here $r$ is the radial coordinate and $\phi$ is the azimuth angle around the origin. 
The equations describing the flow are

\begin{equation}
   \frac{\partial\Sigma}{\partial t}+ \mathbf{\nabla}\cdot (\Sigma\mathbf{v})=0,
	\label{eq:hydrodynamics}
\end{equation}

\begin{equation}
   \frac{\partial \Sigma\mathbf{v}}{\partial t}+ \mathbf{\nabla} \cdot(\Sigma\mathbf{v}\mathbf{v})=-\mathbf{\nabla}P-\Sigma\mathbf{\nabla}\Phi+\mathbf{f_{\nu}},
	\label{eq:hydrodynamics1}
\end{equation}
where $\Sigma$ is the surface density, $\mathbf{v}$ the velocity, $\mathbf{f_{\nu}}$ 
represents the viscous force per unit area and $\Phi$ is the gravitational potential 
in the disc given by
\begin{equation}
\Phi=-\frac{GM_{\star}}{|\mathbf{r}-\mathbf{r}_{\star}|}-\frac{GM_J}{\sqrt{|\mathbf{r}-\mathbf{r}_{J}|^2+\epsilon^2}}-\frac{GM_S}{\sqrt{|\mathbf{r}-\mathbf{r}_S|^2+\epsilon^2}},
 \label{eq:pot}
\end{equation}
where the subscripts $\star$, $J$ and $S$ refer to the central star,
Jupiter and Saturn, respectively. 
The parameter $\epsilon$ is a softening length used to mitigate
computational problems arising from the divergence of the potential 
in the vicinity of the planets. Although we model a regime in which 
the positions of the peaks of the torque density distribution (due 
to the planets) may depend on both $H$ and the planet's Hill radius 
$R_{H}$ \citep[][]{Gennaro2008,Gennaro2010}, we simply use 
$\epsilon= 0.7H_0$, where $H_0$ is the disc scale-height at $r = r_0$
(where $r_0=5.2$~AU is a characteristic radius). 
This value for $\epsilon$ has been applied previously in similar 
contexts \citep[see, e.g.,][]{Morbidellietal2007}. 
In order to gauge the impact of the softening length, a model was also
simulated by applying $\epsilon= 0.5H_0$ (see Appendix~\ref{sec:C}).
In addition, in those models that consider a fixed mass for Saturn 
of $1.8\times10^{-4}M_{\odot}$ 
\citep[about $2/3$ of its final mass, see][]{Walshetal2011},
we apply a cut-off in the calculation of the torque at a distance $R_{H}$
from the planets, to avoid spurious effects in the value of the total
torques acting on the planets (see Appendix~\ref{sec:B}).

We use the version of the hydrodynamic code FARGO, which uses the orbital advection algorithm 
of \citet{Masset2000}, referred to as FARGO2D1D \citep{Cridaetal2007}. This code was implemented with the aim of 
adequately describing the viscous evolution of the disc, which can be important for studying 
the migration of massive planets \citep{Lin1986,Lin1993,Nelsonetal2000,Cridaetal2006,Cridaetal2007,Gennaro2008}. 
FARGO2D1D can account for the long-term evolution of the disc by means of a one-dimensional mesh 
extension, attached to the conventional two-dimensional mesh \citep[see ][for details]{Cridaetal2007}.

In order to assess the reliability of the simulations with FARGO2D1D, we have carried out the experiments in \citet{MassetSnellgrove2001} using the FARGO2D and FARGO3D codes. 
A comparison of the results can be found in Appendix ~\ref{sec:B}. 

Our two-dimensional grid covers a ring with a radial domain from $0.26$ AU to $8.65$ AU, using a linear spacing, with a resolution of $(N_{r},N_{\phi})=(512,384)$ 
zones in radius and azimuth, respectively. The one-dimensional grid extends from $0.026$ AU 
to $36.4$ AU. The initial surface density profile follows a power law, 
$\Sigma(r)\propto r^{-\beta}$, with $\beta=1/2$, which is the value derived for the solar nebula 
by \citet{Davis2005}
(see also discussion in Appendix~\ref{sec:C})
. 
At $r_0=5.2$~AU, the unperturbed surface density, $\Sigma_0$, takes a value 
of $2.25\times10^{-5} M_{\odot}/$AU$^2$, corresponding to $200$~g~cm$^{-2}$. 
The disc is assumed to be locally isothermal, where the (vertically-integrated) pressure is given 
by
\begin{equation}
       P=\Sigma c_s^2.
	\label{eq:pressure}
\end{equation}
The locally-isothermal sound speed, $c_s$, is such that the aspect ratio is constant throughout
the disc, $h=0.05$. 
We use the $\alpha$-prescription \citep[][]{SakuraSunyaev1973} for the kinematic viscosity, 
in which $\nu=\alpha c_s H$. In these calculations, we adopt a value for the turbulence parameter 
$\alpha=10^{-3}$.
A test with a larger value, $\alpha=10^{-2}$, is discussed in Appendix~\ref{sec:C}.

\section{RESULTS}
\label{sec:results}

In this section we present the results of a series of hydrodynamics 
simulations aimed at assessing the effect of the initial 
value of the orbital separation, $\Delta a_{SJ}$, on the orbital evolution of Jupiter and Saturn (see Table~\ref{tab:simulations1}). The initial orbital radii of Jupiter $r_J$ and Saturn
$r_{S}$ were chosen based on the Grand Tack scenario; it is
assumed that Jupiter has already formed (i.e., attained its final mass) and migrated inwards 
to $1.5$~AU, at which point Saturn is assumed to have reached a mass of $\approx 2/3$ of its current value (Runs 1 to 10). 
At that point, Saturn is 
assumed to be near or beyond the ice line, and begins a phase of rapid inward migration
until resonance capture ensues. The initial locations considered for Saturn span a range of 
separations from Jupiter that include many strong mean motion resonances, e.g., the 1:2 and 2:3.
Note that the first two models have Saturn interior of the 1:2 MMR with Jupiter 
(see Table~\ref{tab:simulations1}).

Runs 4a to 7b in Table~\ref{tab:simulations1} were carried out to explore the effect of the mass of Saturn on the resonance capture process and on the direction of migration. A discussion on the role of the different mass history of Saturn is given in 
section~\ref{sec:acc}. 

 \begin{table}
	\centering
	\caption{Parameters for the planets Jupiter and Saturn varied in the simulations. Column 1: Model number. Column 2: Saturn's mass. Column 3: Initial
    orbital radius of Jupiter $r_J$. Column 4: Initial orbital radius of Saturn $r_S$. Column 5: Initial ratio of Keplerian orbital periods between Jupiter and Saturn. Column 6: Initial value for $\Delta a_{SJ}=r_S-r_J$. In all the simulations the mass of Jupiter is $10^{-3}M_{\odot}$.
}
    \begin{threeparttable}[b]
	\label{tab:simulations1}
	\begin{tabular}{|c|c|c|c|c|c|} 
	      \hline
	      \hline
              Run  & $M_S$ &  $r_J$ & $r_S$ &$T_J/T_S$ & $\Delta a_{SJ}$ \\ 
                 &$[M_{\odot}]$&$[AU]$ &$[AU]$& &$[AU]$\\ 
		\hline
		\hline
		1  & $1.8\times10^{-4}$ & 1.5 & 2.0 &1.5396&0.5\\
		2 &$1.8\times10^{-4}$&1.5& 2.23 &1.8126&0.73\\
		3 &$1.8\times10^{-4}$&1.5& 2.5 &2.1516&1.0\\
		4 &$1.8\times10^{-4}$&1.5& 3.0 &2.8284&1.5\\
		5 &$1.8\times10^{-4}$&1.5& 3.5 &3.5642&2.0\\
		6 &$1.8\times10^{-4}$&1.5& 4.0 &4.3546&2.5\\ 
		7 &$1.8\times10^{-4}$&1.5& 4.5 &5.1961&3.0\\
		8 &$1.8\times10^{-4}$&1.5& 5.0 &6.0858&3.5\\
		9 &$1.8\times10^{-4}$&1.5& 5.2 &6.4545&3.7\\
		10 &$1.8\times10^{-4}$&2.0& 4.5 &3.3750&2.5\\
        4a & Accreting\tnote{*}&1.5& 3.0 &2.8284&1.5\\
		7a & Accreting\tnote{*}&1.5& 4.5 &5.1961&3.0\\
        1b & $2.9\times10^{-4}$ & 1.5 & 2.0 &1.5396&0.5\\
		4b &$2.9\times10^{-4}$&1.5& 3.0 &2.8284&1.5\\
		7b &$2.9\times10^{-4}$&1.5& 4.5 &5.1961&3.0\\
		\hline
		\hline
	\end{tabular}
    \begin{tablenotes}
     \item[*] {Saturn is allowed to accrete gas only when it is free to migrate. The initial mass of Saturn is $1.8\times10^{-4}M_{\odot}$ and accretion stops when its mass reaches $2.9\times10^{-4}M_{\odot}$.}
   \end{tablenotes}
  \end{threeparttable}
\end{table}

In the simulations, the orbits of the planets are fixed during the first $2200$ years 
($\approx 200$ orbits at $r_0$), and then the planets are allowed to migrate for 
$\approx 50000$ years, or more (except for models 1b and 4b). 
The phase during which the planets are kept at a fixed orbital distance is aimed 
at allowing Jupiter to open a gap in the gaseous disc, as if it had been undergoing
type~II migration, and Saturn to open a partial gap.
This delay also ensures that the gaps are in quasi-equilibrium when the planets
begin to migrate.

In the following, we adopt an empirical definition for the type of migration after the pair 
of planets is captured in a resonance. If the semi-major axes of Jupiter, when it is caught 
in resonance and at the end of the calculation, differ by $5$\% or more, the migration of 
the pair is classified as either ``Inward'' or ``Outward'' (depending on the sign of the 
difference). Otherwise, the migration is said to be ``Stalled'' (see Table~\ref{tab:simulations2}).
Clearly, this classification may not apply over a longer-term evolution.
We aim to find the values of $\Delta a_{JS}$ for which outward 
migration is feasible and to determine the MMR in which the planets 
become trapped.

\begin{table}
	\centering
	\caption{Summary of results for the simulations. Column 1: Model number. 
	Column 2: Initial value for $\Delta a_{SJ}=r_S-r_J$. Column 3: Saturn's mass. 
	Column 4: Migration direction. Column 5: Final resonance.
	In model~10, capture occurs in the 1:2 MMR, but the pair breaks through
	the resonance because of eccentricity excitation.
    }
	\label{tab:simulations2}
	\begin{tabular}{|c|c|c|c|c|} 
	      \hline
	      \hline
              Run  & $\Delta a_{SJ}$& Saturn's mass & Migration & Final MMR  \\ 
                   &$[AU]$ & $[M_{\odot}]$     \\ 
		\hline
		\hline
		1 & 0.5 &$1.8\times10^{-4}$ & Outwards & 2:3 \\
        2 & 0.73 &$1.8\times10^{-4}$ & Outwards & 2:3 \\
        3 & 1.0 &$1.8\times10^{-4}$ & Outwards & 2:3 \\
		4 & 1.5 &$1.8\times10^{-4}$ & Stalled & 1:2 \\
        5 & 2.0 &$1.8\times10^{-4}$ & Stalled & 1:2 \\
        6 & 2.5 &$1.8\times10^{-4}$ & Stalled & 1:2 \\
		7 & 3.0 &$1.8\times10^{-4}$ & Inwards & 1:2 \\
        8 & 3.5 &$1.8\times10^{-4}$ & Inwards & 1:2 \\
        9 & 3.7 &$1.8\times10^{-4}$ & Inwards & 1:2 \\
        10 & 2.5 &$1.8\times10^{-4}$ & Outwards & 2:3 \\
        4a & 1.5 &Accreting & Stalled & 1:2 \\
		7a & 3.0 &Accreting& Stalled & 1:2 \\
		1b & 0.5 &$2.9\times10^{-4}$ & Outwards & 2:3\\
		4b & 1.5 &$2.9\times10^{-4}$ & Outwards & 2:3 \\
		7b & 3.0 &$2.9\times10^{-4}$ & Outwards & 2:3 \\ 
		\hline
		\hline
	\end{tabular}
\end{table}

Several models considered here produce results similar to others. In fact, referring 
to Table~\ref{tab:simulations1}, we found that the orbital evolution of model~2 is 
quite similar to that of model~1; models~5, 6, 10 evolve similarly to model~4; 
models~8 and 9 show essentially the same behaviour as model~7. 
Therefore, hereafter we focus on models~1, 3, 4, 7, and their
variations 4a, 7a, 1b, 4b and 7b. 
A summary of the resulting MMR and migration direction after capture is reported 
in Table~\ref{tab:simulations2}.

\subsection{Initial migration phase}
\label{sec:initial_mig}

\begin{figure*}
    \includegraphics[width=\linewidth]{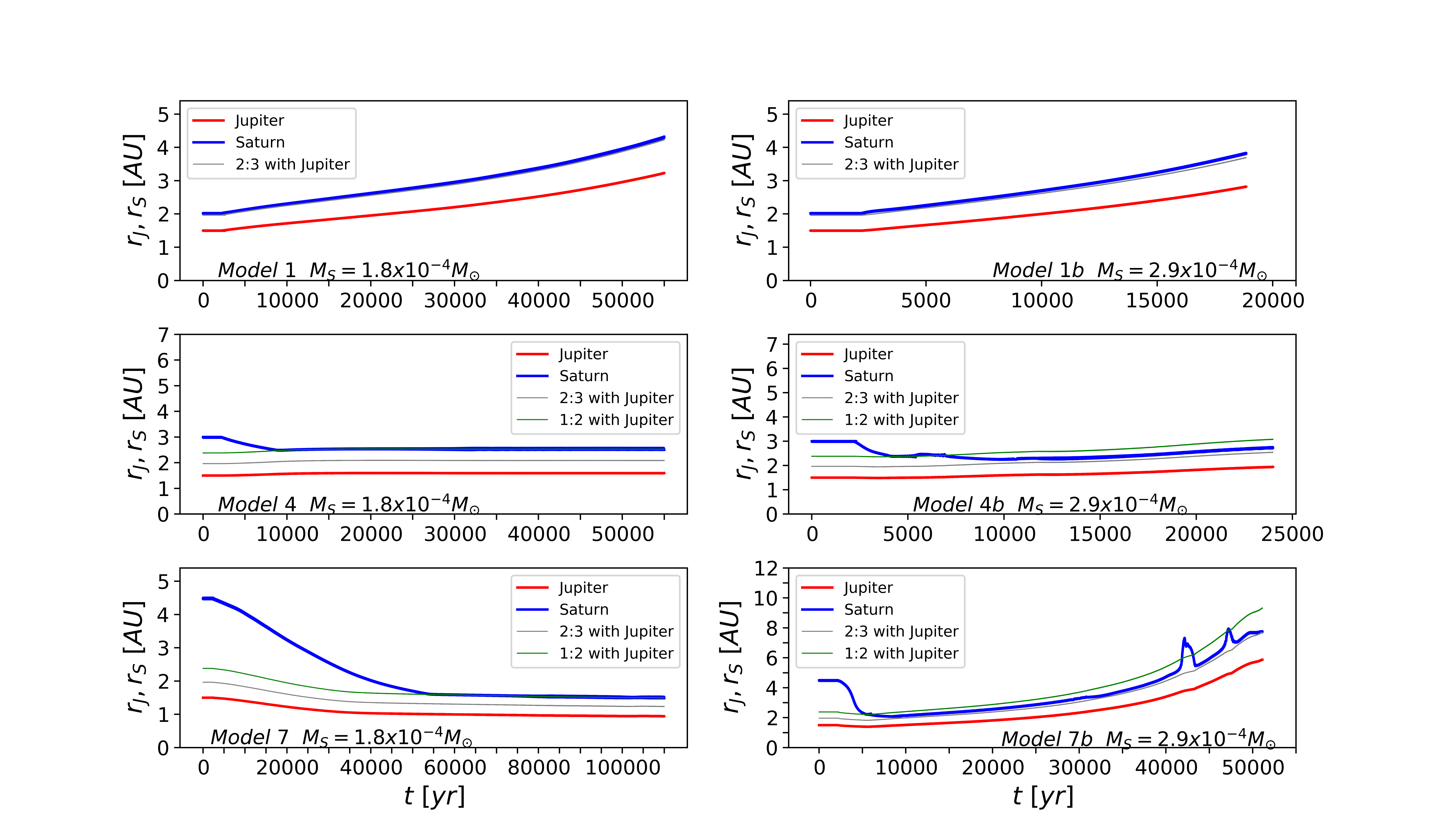}
    \caption{Semi-major axis of Jupiter and Saturn for several models (see Table~\ref{tab:simulations1}). The green and grey curves represent the nominal position of the resonances 1:2 and 2:3 with Jupiter.}
    \label{fig:semi}
\end{figure*}
Figure~\ref{fig:semi} shows the evolution of the orbital radii of Jupiter and Saturn 
in models 1, 4 and 7. 
In model 1, the pair is readily trapped in the 2:3 MMR and starts migrating outwards. 
In models 4 and 7, Saturn is outside of the 1:2 MMR with Jupiter and becomes 
trapped in this resonance as it approaches Jupiter. 
We wish to analyse whether the initial migration of Saturn in 
models 4 and 7 is runaway (i.e., at an ever increasing rate) or a modified type I.

Suppose that the migration of Saturn is type I. If so, the total torque 
(Lindblad plus corotation), $\Gamma$, acting on a planet was estimated by 
\citet{Tanaka2002}. For an isothermal, 2D disc,
\begin{equation}
       \Gamma=-(1.160+2.828\beta)\left(\frac{M_p}{M_\star}\right)^2 h^{-2}\Sigma_pr_p^4\Omega_p^2,
	\label{eq:Tanaka2002}
\end{equation}
where $r_p$ and $\Omega_p$ are the orbital radius and the angular velocity of the planet,
respectively. Here $\Sigma_{p}$ is the surface density at the planet's position.
From equation~(\ref{eq:Tanaka2002}), the type~I radial migration speed of the planet 
can be calculated from the conservation of angular momentum, as
\begin{equation}
       \frac{dr_p}{dt}=2r_p\frac{\Gamma}{L_p},
	\label{eq:Tanaka2002a}
\end{equation}
where $L_p=M_p\sqrt[]{GM_{\star}r_p}$, is the planet's angular momentum. Using
Equations~(\ref{eq:Tanaka2002}) and (\ref{eq:Tanaka2002a}), we can estimate the
migration rates of Saturn for models 4, 7, 4b and 7b, before the planet is caught 
in resonance with Jupiter, and compare them with the actual migration rates of Saturn 
obtained from the calculations. Results are presented in Table~\ref{tab:migration}.
The values in the forth and fifth columns are computed midway between 
the initial orbital radius of Saturn and its capture radius.
The type~I rate is larger than, or comparable to, $\abs{dr_S/dt}$ 
in the models, which suggests that Saturn does not experience a runaway 
migration mode before attaining a resonance with Jupiter.

\begin{table}
	\centering
	\caption{Migration rate of Saturn before becoming locked in resonance with Jupiter. 
	Column 1: Model number. Column 2: Initial value of $\Delta a_{SJ}=r_S-r_J$. 
	Column 3: Saturn's mass. Column 4: Saturn's migration rate. Column 5: Type I migration rate.
	Migration rates are estimated as averages over one orbital period.
    }
	\label{tab:migration}
	\begin{tabular}{|c|c|c|c|c|} 
	      \hline
	      \hline
              Run  & $\Delta a_{SJ}$& Saturn's mass & $\abs{\frac{dr_S}{dt}}$ model& $\abs{\frac{dr_S}{dt}}$  type I \\ 
                   &$[AU]$ & $[M_{\odot}]$  &$[AU/yr]$ & $[AU/yr]$    \\ 
		\hline
		\hline
		1 & 0.5 &$1.8\times10^{-4}$ & $2.14\times10^{-4}$ & $3.30\times10^{-4}$ \\
		4 & 1.5 &$1.8\times10^{-4}$ & $1.00\times10^{-4}$ & $3.17\times10^{-4}$ \\
		7 & 3.0 &$1.8\times10^{-4}$ &$2.21\times10^{-4}$ & $3.00\times10^{-4}$ \\
        4a & 1.5 &Accreting & $1.56\times10^{-4}$ & $4.56\times10^{-4}$ \\
        7a & 3.0 &Accreting & $1.14\times10^{-4}$ & $5.52\times10^{-4}$ \\
		4b & 1.5 &$2.9\times10^{-4}$ & $3.53\times10^{-4}$ & $5.06\times10^{-4}$ \\
		7b & 3.0 &$2.9\times10^{-4}$ & $6.31\times10^{-4}$ & $5.90\times10^{-4}$ \\ 
		\hline
		\hline
	\end{tabular}
\end{table}  

According to equations~(\ref{eq:Tanaka2002}) and (\ref{eq:Tanaka2002a}),
$|dr_{p}/dt|\propto r_{p}^{3/2-\beta}$, thus the migration velocity reduces 
as the planet moves inward (for $\beta=1/2$). If the (convergent) relative 
velocity is dictated by the migration rate of the outer planet, 
compact resonant orbits are more likely obtained farther away from the 
star, where transiting the 1:2 MMR is also more likely \citep{Mustill2011}. 

A condition to initiate runaway migration requires that the surface
density around the planet's orbit be \citep{Gennaro2008}
\begin{equation}
    \Sigma_p\gtrsim h^{2}\left(\frac{M_{\star}}{r^2_p}\right).
	\label{eq:DL08}
\end{equation}
In this case, a strong asymmetry arises in the coorbital streamlines between the leading
and trailing sides of the planet \citep{Ogilvie2006,Gennaro2008}.
This condition, however, is not satisfied in any of the models considered here, as it
would require an unperturbed gas density at $5.2$~AU several times as large as that 
in the calculations. Therefore, runaway migration is not expected according to this
criterion.

According to \citet{Masset2003}, runaway migration can occur after the formation of 
a partial gap, which can result in a coorbital mass deficit, $\delta m$. 
If $\delta m > m_p$, where $m_p=M_p+M_R$, the planet mass augmented by the Roche lobe 
mass, the planet may undergo runaway migration. 
It should be emphasised that the runaway migration domain is highly dependent on how
the coorbital mass deficit is established. We calculated $\delta m$ for Saturn applying 
the expression 
\begin{equation}
       \delta m=2\pi\int_{r_p-x_s}^{r_p+x_s}\left(\Sigma_{p,0}-\Sigma\right) r dr, 
	\label{eq:delta_mass}
\end{equation}
where $x_s$ is half-width of the horseshoe region for 2D discs,
given by \citep{Paardek}
\begin{equation}
       x_s=2.47R_{H},
	\label{eq:x_s}
\end{equation}
and $\Sigma_{p,0}$ is the unperturbed surface density at $r_p$.

\begin{figure}
\includegraphics[width=\linewidth,height=12.5cm]{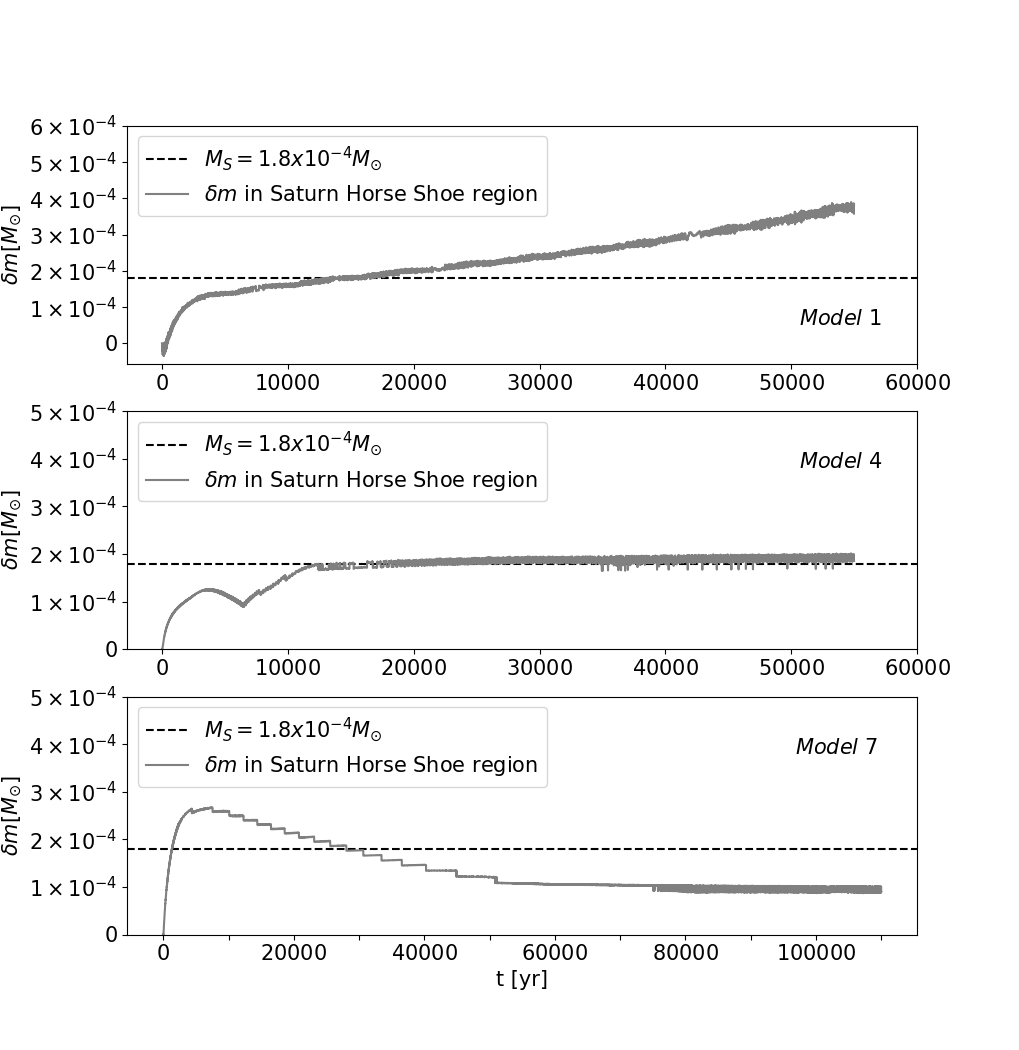}
    \caption{Coorbital mass deficit $\delta m$, in the horseshoe region of Saturn (solid line) in models 1, 4 and 7. The mass of the planet $M_{p}$ is indicated with horizontal lines.}
\label{fig:delta_mass}
\end{figure}
Results of the calculation of $\delta m$ are shown in Figure~\ref{fig:delta_mass}.
Since $M_{R}\ll M_{p}$ during the evolution, the approximation $m_{p}\simeq M_{p}$
holds in the models shown in the Figure. In model 1, Saturn's coorbital 
mass deficit increases with time. However, $\delta m < m_p$ approximately until 
$15000$~yr. 
This suggests that Saturn does not undergo runaway migration prior to resonance capture. 
In model 4, $\delta m$ in Saturn coorbital region first increases and then stays at 
a constant value below $m_p$ (while the planet is captured in resonance with Jupiter, 
see next section). Therefore, runaway migration of Saturn is not expected in this
case either. 
In model 7, $\delta m \gtrsim m_{p}$ during the first $\approx 30000$~yr and 
becomes smaller afterwards. However, even when $\delta m$ reaches its maximum value 
(at $\approx 10000$~yr), $\abs{dr_S/dt}$ in the model is smaller than the rate 
predicted by equations~(\ref{eq:Tanaka2002}) and (\ref{eq:Tanaka2002a}).
Moreover, $\abs{dr_S/dt}$ does not increase during the inward migration phase. 
In fact, it tends to decrease, as expected in type~I migration (as discussion above). 

In summary, all three considerations above agree with the simulations' outcome 
that, if Saturn originates outside the 1:2 MMR, it approaches Jupiter at a type~I rate, 
possibly modified by the perturbed surface density of the disc. The slower radial migration 
of Saturn allows the planets to get trapped in the 1:2 MMR \citep[e.g.,][]{Mustill2011}. 
Afterwards, the migration of the pair is either (nearly) stalled or continues inwards 
(see models 4 and 7 in Figure~\ref{fig:semi}).

\subsection{The role of gap formation}

The orbital evolution of Jupiter and Saturn embedded in a disc is 
complicated because the gaps can overlap 
when the planets have close enough orbital radii. 
Each planet will tend to keep its gap (relatively) clean, 
but the interaction of overlapping gaps can result in 
a complex flow in the surrounding region
\citep[][]{MassetSnellgrove2001,Pierens2011,Gennaro2012,Kley2012}. 
In some cases, this interaction may alter the torques 
exerted on the inner planet, which can result in the outward 
migration of the pair \citep{MassetSnellgrove2001,MorbidelliCrida2007}.

\begin{figure*}
    \includegraphics[scale=1.0]{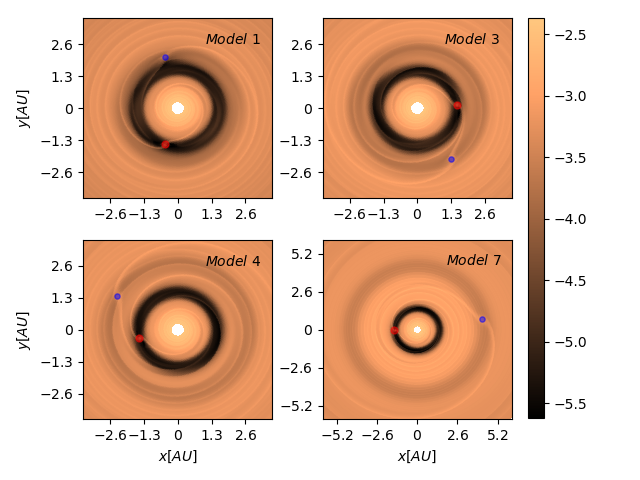}
    \caption{Colour-scale rendering of the surface density 
    in the disc, after $5500$ years of evolution, for models 
    1, 3, 4 and 7, as indicated 
    (see Table~\ref{tab:simulations1} for details). 
    The units of density are
    $M_{\odot}r^{-2}_{0}$.     
    Red and blue circles indicate the position of Jupiter 
    and Saturn, respectively.
    At this time, the planets are
    in resonance in models 1 and 3.
    Capture in MMR occurs at later times in the other two models.
    }
    \label{fig:dens1}
\end{figure*}
Figure~\ref{fig:dens1} shows the disc surface density at $t=5500$~yr, when Jupiter
and Saturn have already opened their respective gaps, and
started their radial migration. In models 1 and 3
($\Delta a_{SJ}\leq 1$~AU, recall that $\Delta a_{SJ}$ is
independent of time and
refers to the initial orbital separation of the pair), Saturn has 
already approached Jupiter close enough so that gaps overlap.
In model 1, Saturn's gap overlaps with the deeper parts of
Jupiter's gap.
In models 1 and 3, the pair is already in MMR at $t\approx 5500$~yr, 
whereas the planets become captured in resonance at later times in 
models 4 and 7 (at $t\approx 8800$ and $\approx 55000$~yr, respectively)  

In model 4 ($\Delta a_{SJ}=1.5$~AU), although the gaps of
Jupiter and Saturn are well separated, the internal
wake of Saturn extends in the gap opened by Jupiter, 
generating a local density enhancement. 
In model 7 ($\Delta a_{SJ}=3.0$~AU), the planets are still
far apart. The gap of Jupiter is basically unaffected by
Saturn's tidal field and \textit{vice versa}. 
The evolution of both planets thus proceeds similarly to that of 
individual bodies interacting with the gaseous disc. 

\begin{figure*}
    \includegraphics[scale=1.0]{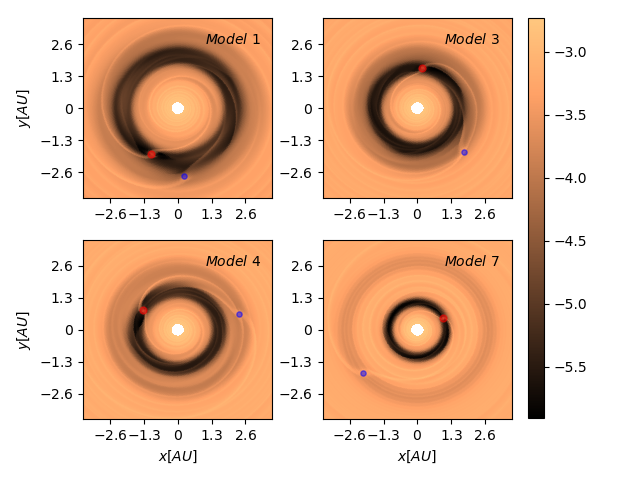}
    \caption{Colour-scale rendering of the surface density
    in the disc, at $t=27500$ years, for models 
    1, 3, 4 and 7, as indicated 
    (see Table~\ref{tab:simulations1}).
    The units of density are
    $M_{\odot}r^{-2}_{0}$.    
    In each case, the red circle indicates
    the position of Jupiter, and the blue circle 
    indicates the location of Saturn.
    At this time, the planets are
    in MMR in models 1, 3, and 4.
    }
    \label{fig:dens}
\end{figure*}
The different density structure in the overlapping 
gap region can lead to a different orbital evolution 
\citep[e.g.,][]{Gennaro2012}.
Figure~\ref{fig:dens} shows the disc surface density for 
the same models as in Figure~\ref{fig:dens1}, but at 
$t=27500$~yr. The long-term evolution of the planets'
orbits is clearly visible in the Figure. The orbital 
radius of Jupiter gradually decreases in going from 
model 1 to model 7. In fact, after capture in
MMR, the planets migrate outward in models 1 and 3, 
their orbits become nearly stationary in model 4, and 
they migrate inward in model 7.

\begin{figure*}
    \includegraphics[scale=1.0]{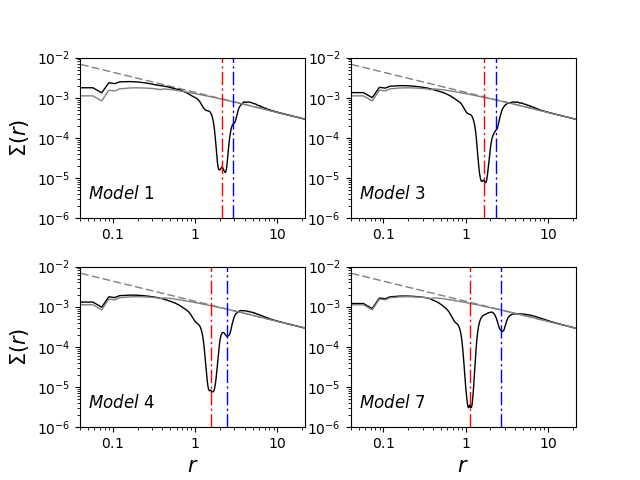}
    \caption{Azimuthally-averaged surface density 
    profiles for models 1, 3, 4 and 7 
    (solid black curve, as indicated, see also
    Table~\ref{tab:simulations1}),
    at $t=27500$ years.
    The units of density are
    $M_{\odot}r^{-2}_{0}$; the radius is in AU.
    The grey dashed lines represent the surface density
    profile at $t=0$, and the grey solid curves correspond
    to the surface density profile in the same disc with
    no planets.
    The red and blue vertical dashed lines indicate the
    radial position of Jupiter and Saturn, respectively.
    The planets are
    in MMR in models 1, 3, and 4, but not yet in model 7.
    }
\label{fig:profiles}
\end{figure*}
Figure~\ref{fig:profiles} shows the azimuthally-averaged 
surface density profiles at $t=27500$~yr. In models 1 and 3, 
Saturn orbits at the outer edge of Jupiter's gap 
(they will eventually become locked in a 2:3 mean motion 
resonance).
In model 4, the gaps of Jupiter and Saturn partially overlap. 
In model 7, Saturn has approached Jupiter, but not close enough 
for their respective gaps to interact and thus their evolution 
is similar to that of isolated planets undergoing type~II 
migration and modified type~I migration, respectively. 
Since gaps in the these simulations are not very deep, Lindblad 
torques are never completely shut off. 

\begin{figure}
    \includegraphics[width=\linewidth]{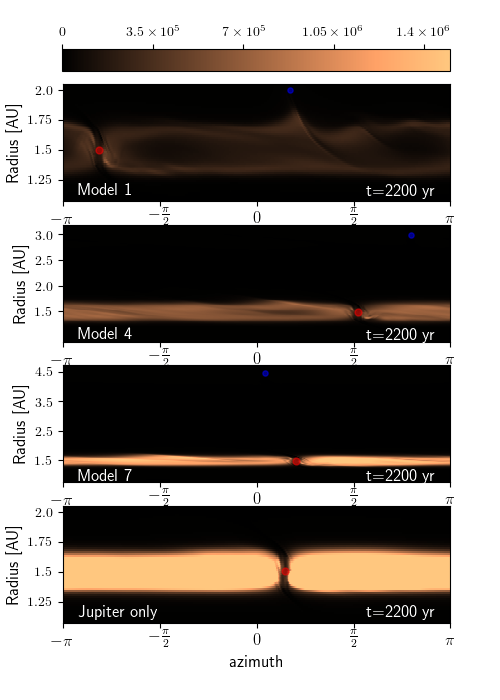}
    \caption{Distribution of vortensity in the coorbital 
    region of Jupiter in the models~1, 4 and 7, as indicated, and in a model
    comprising only Jupiter (bottom panel). The red circle indicates
    the position of Jupiter, whereas the blue circle 
    indicates the location of Saturn.}
\label{fig:vort}
\end{figure}
It is informative to look at the vortensity distribution in the coorbital
region of Jupiter, at the time when the planets start to migrate
($t=2200$ yr, see Figure ~\ref{fig:vort}). This Figure also shows an additional model 
in which only Jupiter is considered. In this case (bottom panel), the distribution of
vortensity in the coorbital region is uniform, indicating (near) saturation of 
the corotation torque on Jupiter.
In models~4 and 7, the vortensity distribution is almost uniform, suggesting that 
the corotation torque is about to saturate. In particular, the vortensity field
in model~7 is quite similar to that of the single-planet case in the bottom panel, 
and therefore Jupiter's migration is nearly unperturbed by Saturn's gravity.
In contrast,
model~1 shows a non-uniform vortensity distribution in Jupiter's coorbital region, 
which is significantly affected by the tidal field of Saturn, as also indicated 
by the strongly perturbed Jupiter's outer gap edge 
(see Figures~\ref{fig:dens1} and \ref{fig:dens}).

\subsection{The role of $\Delta a_{SJ}$ on the direction of migration}

We have explored various initial separations of the planets (see Tables~\ref{tab:simulations1} 
and \ref{tab:simulations2}). In the range of values adopted here, the initial orbital radius of Jupiter 
is not very important because the migration of Saturn is significantly faster.
For lower disc surface densities, though, the situation is expected to be different since 
the relative migration velocity would reduce.

In the Grand Tack model, it is required that the planets migrate outwards once caught
in resonance. This is likely to occur if Saturn orbits within (or close to) the 1:2 MMR
with Jupiter at some point of its evolution, when 
$M_{S}\gtrsim 2\times 10^{-4} M_{\odot}$ \citep[e.g.][]{Gennaro2012}.
In fact, in these situations, convergent migration most likely leads to capture in the
2:3 MMR.
Instead, if Saturn is exterior to the 1:2 resonance, the planets are very likely 
to get trapped in such resonance and outward migration is then unlikely 
\citep[see also][]{Gennaro2012,Pierens2014}.

In model~1, the planets are initially close to the 3:2 MMR ($\Delta a_{SJ}=0.5$~AU, see Figure~\ref{fig:semi}). 
Because of the orbits' proximity, the disc region around Jupiter's gap (outside its orbit) 
is affected by the tidal field of Saturn (see Figure~\ref{fig:dens1}). When Saturn transits 
the 2:3 MMR, the exterior (negative) Lindblad torque exerted by the disc on Jupiter is weakened 
by the presence of Saturn's gap (i.e., the torque is reduced due to the lower density), and 
the interior (positive) Lindblad torque can tip the overall torque balance, driving the inner 
planet (and hence the pair) outwards. The situation in model~2 is similar 
($\Delta a_{SJ}=0.73$~AU). In model~3 ($\Delta a_{SJ}=1.0$~AU), Saturn starts just outside 
the 1:2 MMR with Jupiter but when it transits the resonance, there is not gap overlap. 
The migration velocity is large enough to overcome the forcing by the resonance and,
therefore, Saturn continues its inward migration until it reaches the 2:3 MMR.

In principle, if the migration rate of the interior planet is much smaller than 
that of the exterior planet and the latter migrates at a type~I rate, what determines 
resonance capture is the orbital radius at which resonances are crossed 
(assuming orbital eccentricity can be neglected). This is because the relative 
migration velocity of the exterior planet depends on $r_{p}$, assuming a power-law 
disc, and is approximately $\propto r_{p}$ in these simulations 
(see section \ref{sec:initial_mig}).
If the initial semi-major axis of Jupiter is the same, as in models~4 to 9,
but the initial semi-major axis of Saturn progressively increases (from $3$ in model~4 
to $5.2$~AU in model~9, see Table~\ref{tab:simulations1}), the 1:2 MMR is crossed 
at a somewhat shorter orbital radius, because of the inward migration of Jupiter. 
Consequently, as the initial orbital separation $\Delta a_{SJ}$ increases, the relative
velocity at which Saturn crosses the 1:2 MMR with Jupiter reduces in magnitude. 
Since model~4 shows locking in this resonance, all other models are expected to be 
locked in this resonance too (since they approach the resonance at lower relative 
velocity). 
The results from the calculations agree with these arguments (see
\S \ref{tab:simulations2}), including with the fact that the 1:2 MMR is crossed
at smaller radii as $\Delta a_{SJ}$ increases.
It should be pointed out that, if the exterior planet was instead subjected to 
a type~III migration regime (e.g., if the gas density was much larger), the outcome 
would be different and the relative migration velocity would increase (in magnitude) 
with increasing $\Delta a_{SJ}$.

In models~4--9, the dynamical behaviour after resonance capture may depend on some subtle 
differences, like the density perturbations induced by tides in the disc and the somewhat 
different radii at which the resonance is crossed. In some cases (models~4-6), the resulting 
torques acting on Jupiter nearly balance, resulting in a stalled pair (i.e., Jupiter's
semi-major axis varies by less than 5\% after resonance capture). 
In the other cases 
the total torque remains negative, and inward migration continues. Clearly, it is possible 
that the observed behaviour may change over the longer, viscous timescale of the disc.
However, in some models, e.g., in model~7, we do simulate the coupled migration of the planet
pair for a time period comparable to the local viscous timescale.

Model~10 has the same initial separation as model 6, but Jupiter starts at 
an orbital radius of $2$~AU. Therefore, compared to model~6, resonances are crossed 
farther out in the disc. This implies that the relative migration speed at resonance 
crossing is larger (in magnitude) in model~10. Nonetheless, when Saturn transits the
1:2 MMR with Jupiter, it becomes locked in resonance. While the pair evolves in this
resonance, the orbital eccentricity of both planets increases substantially, 
undergoing oscillation cycles \citep[see also, e.g.,][]{Marzari2019}. 
On average, Jupiter's eccentricity grows up to $0.25$
and that of Saturn to $0.15$. But as the amplitude of Jupiter's eccentricity oscillation
increases, so does that of Saturn. Eventually, as the oscillation cycle takes Jupiter's 
eccentricity up to values of $\approx 0.3$, that of Saturn reaches $\approx 0.5$.
The planet skips the resonance and migrates toward Jupiter, until it is locked in the
2:3 MMR. In this resonant configuration, Jupiter's eccentricity damps to small values 
but that of Saturn remains above $0.1$ until the end of the calculation.

The results from models~1--10 suggest that a scenario in which Saturn is captured 
in the 2:3 MMR with Jupiter is most feasible if the planets originate in a compact
configuration, e.g, Saturn is close to or inside the 1:2 MMR, when its mass is
$\approx 1.8\times 10^{-4}\, M_{\odot}$, $\approx 2/3$ of its final mass. 
(Model~10 is peculiar in that capture
is still in the 1:2 MMR, but continuous excitation of orbital eccentricity leads to
further evolution.)
Capture in the 2:3 MMR results in the outward migration of the pair.
Otherwise, if a forming Saturn is farther away from Jupiter, capture occurs in the
1:2 MMR and migration either continues inward or is stalled.

\subsection{The effect of Saturn's mass on migration and resonance capture}
\label{sec:acc}

\begin{figure*}
    \includegraphics[width=\linewidth]{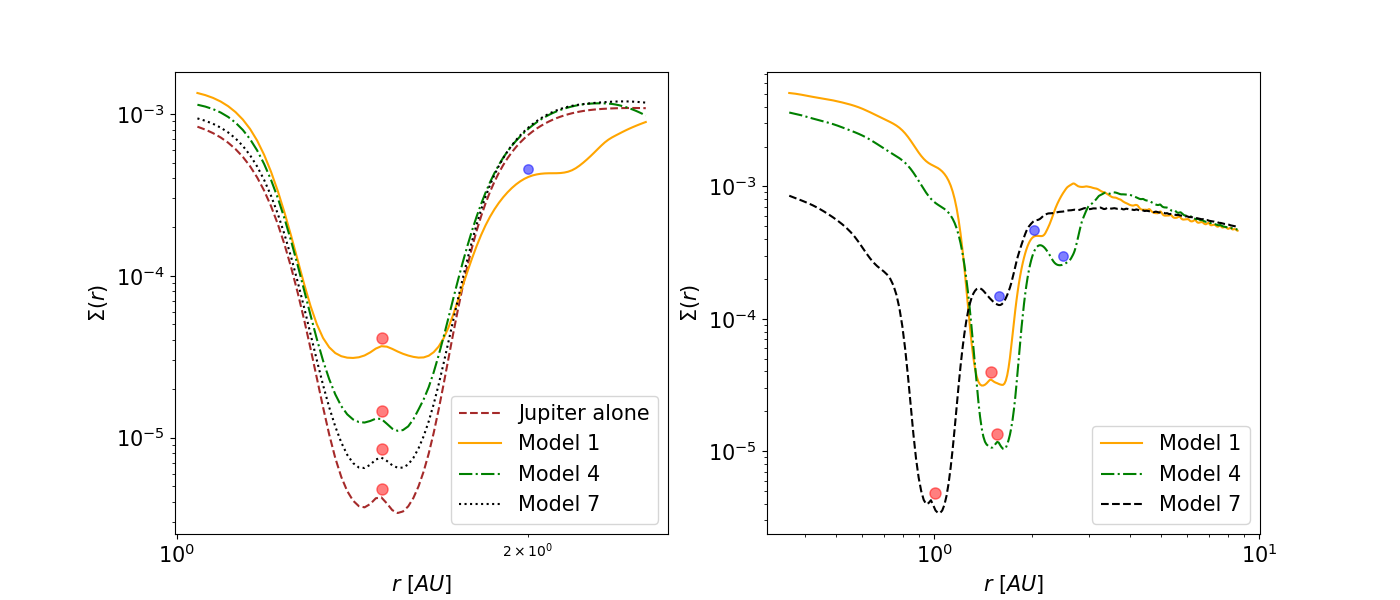}
    \caption{\textit{Left:} Close-up view of the azimuthally-averaged surface density
    profiles in the gap region of Jupiter at $t=2200$ yr. 
    Profiles are shown for the simulations
    with a single Jupiter-mass planet (``Jupiter alone'' case) 
    and for models 1, 4 and 7, as indicated
    (see Table~\ref{tab:simulations1}).\textit{Right:} Azimuthally-averaged surface
    density
    profiles in the gap region of Jupiter for
    the models 1, 4, and 7, at time when the
    two planets are trapped in resonance.
    The units of density are
    $M_{\odot}r^{-2}_{0}$; the radius is in AU. Again, the red circles indicate
    the position of Jupiter, whereas the blue circles 
    indicate the location of Saturn.
    }
\label{fig:jup_gaps}
\end{figure*}
In Table~\ref{tab:simulations2}, we show results from the list of selected simulations,
including variations on models 1, 4, and 7. In addition,
Figure \ref{fig:semi} shows the semi-major axis of Jupiter and Saturn as a function
of time for models 1, 4, 7, 1b, 4b and 7b.
In model 1, 2, and 3, in which the mass of Saturn is fixed at a value of 
$1.8\times 10^{-4}M_{\odot}$, Saturn becomes trapped in the 2:3 MMR with 
Jupiter. The planet pair migrates outward thereafter.
This result is expected because of the assumption that
Saturn originates inside (or very close to) the 1:2 MMR with Jupiter, and
the 2:3 is the next resonance in which
capture is highly likely.
Once in resonance, gap overlap leads to a positive torque exerted on Jupiter.
Instead, in model 4, which has a larger initial separation $\Delta a_{SJ}$, 
Saturn becomes trapped in the 1:2 MMR, as tidal forcing cannot overpower 
resonance forcing. Thereafter, both planets exhibit a stalled migration 
(as defined above), due to a relatively small net torque acting on the planets
(a similar behaviour is observed for models 5 and 6).
In model 7, once Saturn approaches Jupiter and becomes locked in the 1:2 resonance, 
the planets slowly migrate inward (see Figure~\ref{fig:semi}).
The different migration behaviour in models 4 and 7 can be explained in terms 
of the orbital separations and gap overlap after resonance capture. The orbital 
configuration in model 7 favours a more negative total torque exerted on Jupiter, 
hence the inward migration (compare the density distributions for the two cases 
in Figure~\ref{fig:jup_gaps}).
The final outcome of models 8 and 9 resembles that of model 7.

For a fixed mass value of $2.9\times10^{-4}\,M_{\odot}$ for Saturn (corresponding 
to Saturn's current mass), and for $\Delta a_{SJ}$ equal to those considered in 
models 1, 4 and 7 (that is, for models 1b, 4b, and 7b), the migration of the planets 
proceeds outward after resonance capture. 
The outcome of model 1b is similar to that of model 1, as expected. 
The results of models 4b and 7b, however, differ from those of their counterparts, 
models 4 and 7.

In model 4, capture occurs in the 1:2 MMR, whereas it does in 
the 2:3 MMR in model~4b, because the approaching
speed to Jupiter is higher in the latter case (see Table~\ref{tab:migration}). 
Consequently, the two models exhibit different migration directions after resonance 
capture.
The stalled/outward migration obtained in these models can be related to the gap
structures of the two planets.
In model~4, Saturn's gap abates the outer Lindblad torque on Jupiter to the point 
that there is nearly a zero net torque acting on Jupiter (hence the ``stalled'' 
migration). 
Instead, in model~4b, the greater overlap of Saturn's gap (with Jupiter's) produces 
a larger reduction of the outer Lindblad torque (on Jupiter), which is then outweighed 
by the inner Lindblad torque, resulting in a positive total torque acting on Jupiter 
and in the outward migration of the pair.

A similar argument can explain the different outcomes of models 7 and 7b. 
Although the migration rate of Saturn is still within the type~I regime
in both models (see Table~\ref{tab:migration}), the approach velocity of 
Saturn to Jupiter is four times as large in model~7b, and large enough for 
the tidal forcing to exceed the 1:2 MMR forcing. 
Saturn transits the resonance and becomes trapped in the next first-order 
resonance, the 2:3.
Again, because of the more compact orbital architecture and overlapping 
gaps, outward migration is a likely outcome in this configuration
\citep[as also noted in many previous studies, e.g.,][]{MassetSnellgrove2001,MorbidelliCrida2007,Pierens2011,Gennaro2012}.
Henceforth, the results from models~4b and 7b show that it is important 
to examine the accretion history of Saturn prior to capture in resonance 
with Jupiter.

Based on the results presented above, we studied the migration of the planets 
including the effect of gas
accretion on Saturn (models labelled as ``accreting'' in Tables~\ref{tab:simulations1}, 
\ref{tab:simulations2}, and \ref{tab:migration}). 
The role of the accretion on the migration of Jupiter and Saturn was investigated by
\citet{MassetSnellgrove2001} and \citet{MorbidelliCrida2007} in two-dimensional calculations. 
These authors found that accretion does not have a strong impact on the rate and direction of
migration. Both studies used an accretion prescription similar to the one given in \citet{Kley}. 
Since accretion of gas is typically a three-dimensional phenomenon \citep{Gennaro2003},
we apply a formula for the mass accretion of Saturn similar to that derived via high-resolution 
three-dimensional hydrodynamics calculations by \citet{Bodenheimer2013}, which represents 
a disc-limited accretion rate, i.e., the maximum sustained rate of accretion a planet can
achieve. 
We considered an initial mass for Saturn equal to $1.8\times10^{-4}M_{\odot}$ and 
gas accretion is activated once the planet is free to migrate. At this mass, 
Saturn is likely within the runaway accretion phase, although growth is not (yet) 
necessarily limited by the supply rate of the surrounding disc \citep[e.g.,][]{Lissauer2009}. 
Therefore, we may overestimate the gas accretion rate and underestimate the growth 
timescale of the planet. Nonetheless, at this stage of growth, contraction may 
indeed dictate mass-doubling timescales of $\sim 1000$ years 
\citep[][]{Hubickyj2005,Lissauer2009,Movshovitzetal2010}.
Accretion on Jupiter is neglected. Given the large surface density of the disc, 
the accretion rate is relatively large (for the assumed disc conditions) and 
it takes about $600$ years to gain the remainder of the mass and reach 
the final value of $2.9\times10^{-4}M_{\odot}$, at which point accretion is 
assumed to stop.
In reality, however, Saturn would keep growing. In doing so, it would reduce
the inward flow of gas, stalling Jupiter's growth and possibly becoming more 
massive than Jupiter if no intervening effect (such as disk dispersal) stopped
Saturn's accretion. 
The evolution of the pair would then be affected more drastically \citep{Gennaro2012}.

Under these assumptions, the orbital evolution of the pair in models~4a and 7a is not 
significantly affected by the growing mass of Saturn and the results are similar to those 
obtained from models with fixed masses, models~4 and 7 (capture in the 1:2 MMR and
stalled/inward migration, see Table~\ref{tab:simulations2}).
Given the same initial conditions, including gap formation prior to the migration phase, 
this is not surprising and is in accord with the findings of 
\citet{MassetSnellgrove2001} and \citet{MorbidelliCrida2007}.
Overall, within the context of this study,
models~4, 4a, and 4b and models~7, 7a, and 7b indicate that disk conditions 
(e.g., gap formation) when Saturn initiates its migration may impact resonance capture
(i.e., 1:2 versus 2:3 MMR) more than Saturn's final growth history, 
from $2/3$ of the mass 
to full mass. Were Saturn to start migrating at an earlier formation stage, that is, at
a lower mass, it is more unlikely that it would transit across the 1:2 MMR with Jupiter 
(because of the lower relative migration velocity). In this case, starting close or
inside the 1:2 MMR may not help since the migration of the pair after resonance capture 
would be inward \citep{Gennaro2012}.

\section{CONCLUSIONS}
\label{sec:conclusions}

In this paper we present results obtained from 2D hydrodynamics models of the orbital evolution 
of two planets, representing young Jupiter and Saturn, driven into a MMR by convergent migration. 
We analysed in detail the influence in the orbital dynamics of an important parameter, the relative 
initial distance between the planets' (circular) orbits $\Delta a_{SJ}$, in the context of the 
Grand Tack scenario for the evolution of Jupiter and Saturn.
The impact of this parameter has not been previously assessed. 

We found that the inward-then-outward migration proposed in the Grand Tack is also dependent on 
the initial separation of the planets. In most cases, the initial orbital configurations drives 
the planets in a 1:2 MMR (see Table~\ref{tab:simulations2}), after which the pair either migrates 
inward or experiences a stagnant migration (for the duration of the calculations, e.g., model 4). 
Locking in the 2:3 MMR typically requires the pair to start from a compact orbital configuration,
close to or inside the 1:2 MMR.

Among the experiments reported here, Model~$7$ is the one that most closely resembles
the conditions envisioned in the Grand Tack scenario. 
There, Jupiter is supposed to undergo
type~II migration from $3.5$ to $1.5$~AU, over a period of $\sim 10^{5}$ years. During
this time, Saturn is supposed to grow from $30$ to $60\, M_{\oplus}$, in place 
at $4.5$~AU, and then start migrating inward. 
The disc conditions of Model~$7$ are compatible
with both requirements. In fact, the applied gas viscosity would allow Jupiter to
migrate (according to a classical type~II rate, $-3/2\langle\nu/r_{J}\rangle$) 
over a radial distance of $\sim 2$~AU in $\sim 10^{5}$ years.
At a mass of $30\, M_{\oplus}$, Saturn would still be in a thermally-regulated
contraction phase, but the accretion rate would be large enough to gain 
$\sim 30\, M_{\oplus}$ in $10^{5}$ years \citep{Lissauer2009}. 
If accretion became limited by the disc at some point, the gas density would 
be sufficiently large to allow the planet to reach $60\, M_{\oplus}$ in far 
less than $10^{5}$ years \citep{Lissauer2009}.
Assuming, as done in the Grand Tack, that Saturn's growth suddenly slowed down and stopped, 
Model~$7$ reproduces all necessary requirements. Yet, once Saturn reaches the 1:2
MMR with Jupiter, it is locked in resonance and the pair continues to migrate inward.

Models $5$ through $9$ can be considered as variations of Model~$7$ in which 
the initial migration location of Saturn is pushed either inward or outward. 
In all cases resonance locking occurs in the 1:2 MMR and in none the pair reverses 
the course of migration.
Therefore, the outcome we find appears robust against relatively small changes of
the initial conditions.
Models $1$ through $4$ would require some amount of radial migration of Saturn
while it is growing from $30$ to $60\, M_{\oplus}$ (without approaching Jupiter 
too closely).
But in these cases orbital locking occurs in the 2:3 MMR (with subsequent outward
migration) only if Saturn starts (at $2/3$ of its final mass) close to or inside the 1:2 
MMR location of a fully-formed Jupiter.

Other initial conditions, like those used in Model~10 and leading to
more exotic orbital behaviours, are unlikely to offer a viable option, 
since the rise in orbital eccentricity of the two giant planets would 
disrupt (and probably heavily deplete) the asteroid belt. Moreover, 
it remains unknown whether Saturn's eccentricity would damp to current 
values.

We also discuss experiments in which Saturn starts to migrate when it is fully
formed (``b'' models) and by applying a disk-limited accretion rate when migration
starts (``a'' models). Jupiter's mass is always fixed at its final value. 
The accreting models result in a behaviour similar to those of the 
reference models ($M_S=1.8\times10^{-4}M_{\odot}$) whereas the final-mass
models ($M_S=2.9\times10^{-4}M_{\odot}$) tend to establish a 2:3 MMR and 
reverse course of migration. 
However, gravitational perturbations induced by a more massive Saturn 
in the asteroid belt, upon approaching Jupiter, are expected to be more 
disruptive.

We focused here on the late accretion phase of Saturn, neglecting 
ongoing accretion on Jupiter. In reality, both processes should 
be taken into account concurrently. Although the growth timescale 
of Jupiter, $M_{J}/\dot{M}_{J}$, would probably be longer than that of
Saturn, effects on the disk evolution, due to the depletion of the
inner disk can alter the orbital evolution of the pair
\citep{Gennaro2012}. Nonetheless, Saturn's growth is likely to 
determine the type of MMR in which the pair is eventually locked. 
Given the gas surface density required by the scenario to drive 
the pair outward by several astronomical units, the final mass 
of Saturn should be attained relatively quickly compared to 
its migration timescale.
More problematic, however, is to reconcile the lack of growth of
both planets during the reversed migration phase which, in the best
of circumstances, could alter the inner planet mass somewhat, 
but it would significantly change the outer planet mass. 
Nonetheless, Saturn's ``initial'' mass (i.e., when rapid migration 
toward Jupiter begins) and the orbital separation at that time, 
$\Delta a_{SJ}$, are likely to play a fundamental role in the scenario's 
outcome.

 \section*{Acknowledgements}
We thank the reviewer for insightful comments.
The computers Tycho 2 and Geminis (Posgrado en Astrof\'{\i}sica-UNAM, Instituto de Astronom\'{\i}a-UNAM, Instituto de Astronom\'{\i}a Ensenada-UNAM and PNPC-CONACyT) was used to conduct this research. R.O. acknowledges postdoctoral CONACyT grant. This work was partially supported by PAPIIT project IN111118.
G.D. acknowledges support support from NASA's Research Opportunities in Space 
and Earth Science grant 80HQTR19T0071.



\bibliographystyle{mnras}

\begin{thebibliography}{99}


\bibitem[\protect\citeauthoryear{Bodenheimer et al.}{2013}]{Bodenheimer2013} Bodenheimer P., D'Angelo G., Lissauer J. J., Fortney J. J., Didier Saumon 2013, \apj, 770, 120

\bibitem[\protect\citeauthoryear{Ciesla \& Cuzzi}{ 2006}]{CieslaCuzzi2006}Ciesla F., \& Cuzzi J. N. 2006, Icarus, 181, 178


\bibitem[\protect\citeauthoryear{Crida et al.}{ 2006}]{Cridaetal2006}Crida A., Morbidelli A., Masset F. 2006, Icarus, 181, 587

\bibitem[\protect\citeauthoryear{Crida et al.}{ 2007}]{Cridaetal2007}Crida A., Morbidelli A., \& Masset F. 2007, A\&A, 461, 1173

\bibitem[\protect\citeauthoryear{D'Angelo et al.}{2003}]{Gennaro2003}D'Angelo G., Kley W., \& Henning T., 2003, \apj, 586, 540 

\bibitem[\protect\citeauthoryear{D'Angelo \& Lubow}{2008}]{Gennaro2008}D'Angelo G., \& Lubow S.~H., 2008, \apj, 685, 560 

\bibitem[\protect\citeauthoryear{D'Angelo \& Lubow}{2010}]{Gennaro2010}D'Angelo G., \& Lubow S.~H., 2010, \apj, 724, 730 

\bibitem[\protect\citeauthoryear{D'Angelo \& Marzari}{2012}]{Gennaro2012}D'Angelo G., \& Marzari F., 2012, \apj, 757, 50

\bibitem[\protect\citeauthoryear{D\"urmann \& Kley}{2015}]{Duermann2015}D\"urmann C., \& Kley W., 2015, \aap,  574, A52

\bibitem[\protect\citeauthoryear{Duffell et al.}{2014}]{Duffell2014}Duffell P.~C., Haiman Z., MacFadyen A.~I., D'Orazio D.~J., \& Farris B.~D., 2014, \apjl, 792, L10

\bibitem[\protect\citeauthoryear{Davis}{2005}]{Davis2005}Davis S. S., \apj, 627, L153 


\bibitem[\protect\citeauthoryear{Gomes et al.}{2005}]{Gomesetal2005}Gomes R., Levison H. F., Tsiganis K., \& Morbidelli A. 2005, Nature, 435, 466

\bibitem[\protect\citeauthoryear{Hubickyj et al.}{2005}]{Hubickyj2005}
Hubickyj O., Bodenheimer P., \& Lissauer J. J.,  2005, Icarus, 179, 415

\bibitem[\protect\citeauthoryear{Kanagawa et al.}{2018}]{Kanagawa2018}Kanagawa K.~D., Tanaka H., \& Szuszkiewicz E., 2018, \apj, 861, 140 

\bibitem[\protect\citeauthoryear{Kley}{1999}]{Kley} Kley W. 1999, MNRAS, 303, 696

\bibitem[\protect\citeauthoryear{Kley}{2000}]{Kley2000}Kley W. 2000, MNRAS.,
313, L47

\bibitem[\protect\citeauthoryear{Kley et al.}{2005}]{Kleyetal2005}Kley W., Lee, M. H., Murray, N., \& Peale, S. J., 2005, A\&A, 414, 735

\bibitem[\protect\citeauthoryear{Kley \& Nelson}{ 2012}]{Kley2012}Kley W., Nelson R. P. 2012, Annu. Rev. Astron. Astrophys.,
50, 211

\bibitem[\protect\citeauthoryear{Lee \& Peale}{2002}]{LeePeale2002}Lee M. H., \& Peale S. J., 2002, ApJ, 567, 596:609

\bibitem[\protect\citeauthoryear{Levison et al.}{2008}]{Levisonetal2008}Levison H. F., Morbidelli A., Vanlaerhoven C., Gomes R., \& Tsiganis K. 2008, Icarus, 196, 258

\bibitem[\protect\citeauthoryear{Lin \& Papaloizou}{1986}]{Lin1986}Lin D. N. C., \& Papaloizou J. C. B. 1986, \apj, 309, 846

\bibitem[\protect\citeauthoryear{Lin \& Papaloizou}{1993}]{Lin1993}
Lin D. N. C.,  Papaloizou J. C. B. 1993, in Protostars and Planets III, ed. E. H. Levy \& J. I. Lunine (Tuczon, AZ: Univ. Arizona Press), 749-835 

\bibitem[\protect\citeauthoryear{Lissauer et al.}{2009}]{Lissauer2009}
Lissauer J. J., Hubickyj O., D'Angelo G., \& Bodenheimer P., 2009, Icarus, 199, 338

\bibitem[\protect\citeauthoryear{Marzari et al.}{2019}]{Marzari2019}Marzari F., D'Angelo G., \& Picogna G., 2019, \aj, 157, 45

\bibitem[\protect\citeauthoryear{Masset}{2000}]{Masset2000}Masset F. S. 2000, A\&AS, 141, 165


\bibitem[\protect\citeauthoryear{Masset \& Papaloizou}{2003}]{Masset2003}Masset F. S \& Papaloizou J. C. B., 2003, \apj, 588, 494

\bibitem[\protect\citeauthoryear{Masset \& Snellgrove}{2001}]{MassetSnellgrove2001}Masset F. S \& Snellgrove M., 2001, MNRAS, 320, L55

\bibitem[\protect\citeauthoryear{Morbidelli et al.}{2009}]{Morbidellietal2009a}Morbidelli A., Brasser R., Tsiganis K., Gomes R., \& Levison H. 2009, AAS/Division for Planetary Sciences Meeting Abstracts, 41.55.03

\bibitem[\protect\citeauthoryear{Morbidelli \& Crida}{2007}]{MorbidelliCrida2007}Morbidelli A., \& Crida A. 2007, Icarus, 191, 158

\bibitem[\protect\citeauthoryear{Morbidelli et al.}{2005}]{Morbidellietal2005}Morbidelli A., Levison H. F., Tsiganis K., \& Gomes R. 2005, Nature, 435, 462

\bibitem[\protect\citeauthoryear{Morbidelli \& Nesvorn\'y}{2019}]{Morbidelli2019}Morbidelli A.,\& Nesvorn\'y D. 2019, in ``The Transneptunian Solar System'', eds. Dina Prialnik, Maria Antonietta Barucci, \& Leslie Young (Elsevier) 

\bibitem[\protect\citeauthoryear{Morbidelli et al.}{2007}]{Morbidellietal2007}Morbidelli A., Tsiganis K., Crida A., Levison H. F., \& Gomes R. 2007, \apj, 134, 1790

\bibitem[\protect\citeauthoryear{Movshovitz et al.}{2010}]{Movshovitzetal2010}Movshovitz N., Bodenheimer P., Podolak M., \& Lissauer J.~J. 2010, Icarus, 209, 616

\bibitem[\protect\citeauthoryear{Mustill \& Wyatt}{2011}]{Mustill2011}Mustill A.~J., \& Wyatt M.~C. 2011, MNRAS, 413, 554

\bibitem[\protect\citeauthoryear{Nelson et al.}{2000}]{Nelsonetal2000}Nelson R. P., Papaloizou J. C. B., Masset F., \& Kley, W. 2000, MNRAS, 318, 18

\bibitem[\protect\citeauthoryear{Nesvorn\'y}{2018}]{Nesvorny2018}
Nesvorn\'y D. 2018, ARAA, 56, 137

\bibitem[\protect\citeauthoryear{Ogilvie \& Lubow}{2006}]{Ogilvie2006}Ogilvie G. I., \& Lubow S. H., 2006, MNRAS, 370, 784

\bibitem[\protect\citeauthoryear{Paardekooper \& Papaloizou}{2009}]{Paardek} Paardekooper S. J. \& Papaloizou J. C. B., 2009, MNRAS, 394, 2297

\bibitem[\protect\citeauthoryear{Pierens \& Raymond}{2011}]{Pierens2011}Pierens A., \& Raymond, S. N. 2011, A\&A, 533, A131

\bibitem[\protect\citeauthoryear{Pierens et al.}{2014}]{Pierens2014}Pierens A., Raymond S. N., Nesvorny D., Morbidelli, A., 2014, \apjl, 795, L11

\bibitem[\protect\citeauthoryear{Shakura \& Sunyaev}{1973}]{SakuraSunyaev1973}Shakura N. I., \& Sunyaev R. A., 1973, A\&A, 24, 337 

\bibitem[\protect\citeauthoryear{Tanaka et al.}{2002}]{Tanaka2002}Tanaka H., Takeuchi T., Ward W. R. 2002, ApJ, 565, 1257

\bibitem[\protect\citeauthoryear{Tsiganis et al.}{2005}]{Tsiganisetal2005}Tsiganis K., Gomes R., Morbidelli A., \& Levison H. F. 2005, Nature, 435, 466


\bibitem[\protect\citeauthoryear{Walsh et al.}{2011}]{Walshetal2011}Walsh K. J., Morbidelli A., Raymond S. N., O'Brien D. P., \& Mandell A. V. 2011, Nature, 475, 206

\bibitem[\protect\citeauthoryear{Zhang \& Zhou}{2010a}]{ZhangZhou2010}Zhang H., \& Zhou J., 2010, \apj, 714,532


\bibitem[\protect\citeauthoryear{Zhang \& Zhou}{2010b}]{Zhang-Zhou2010b}Zhang H., \& Zhou J., 2010, \apj, 719,671

\end{thebibliography}


\appendix

\section{Variations of model~7}
\label{sec:C}

The gravitational potential in Equation~(\ref{eq:pot}) is singular at the
position of the planets, which can lead to artificially large gravitational
forces exerted on the gas, especially when the planets move through the
grid. The softening length $\epsilon$ is used to smooth the potential
and mitigate the effect of spurious forces. However, this parameter can
affect the torques exerted on the planets by surrounding gas. In order
to asses the impact of the softening parameter, model~7 
($\epsilon=0.7 H_{0}$) was also calculated
by applying $\epsilon=0.5 H_{0}$. Outcomes of these two simulations are
broadly consistent, with Saturn approaching Jupiter and becoming trapped
in the 1:2 MMR.

In this study, we chose the gas viscosity by using a fixed value of $\alpha=10^{-3}$.
Since disc viscosity can affect the gap structure, also migration (hence resonance
capture) can be impacted. If changes in $\alpha$ are limited to factors of several, 
Saturn's evolution is expected to be affected the most.
We performed an additional calculation, equivalent to model~7, but applying a 
higher viscosity parameter, $\alpha=0.01$. In this situation, the shallower gap 
produced by Saturn may enhance its initial migration velocity so that, once reached 
the 1:2 MMR with Jupiter, the planet may be able to overcome the resonance forcing
and proceed inward. In fact, as expected, Saturn's migration is initially larger 
than it is in model~7 and the planet approaches Jupiter in a shorter timescale. 
However, the relative migration velocity is not large enough to allow Saturn 
to break through the 1:2 MMR and, ultimately, the pair shows a stalled migration.
Nonetheless, it should be noted that the orbital migration behaviour could be 
qualitatively different if a much higher or lower viscosity than that applied
here was adopted.

\begin{figure}
\includegraphics[width=\linewidth]{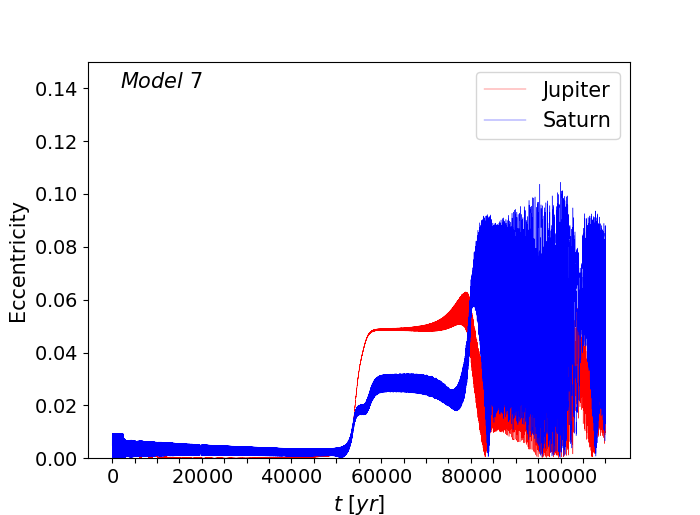}
    \caption{Orbital eccentricity of  Jupiter and Saturn 
    versus time in the model~7.}
\label{fig:ecc_m7}
\end{figure}
Figure \ref{fig:ecc_m7} shows the evolution of the eccentricity of Jupiter and
Saturn in model~7. The 2:1 MMR configuration between the planets causes both
orbital eccentricities to become excited, but they remain relatively small and 
no instability develops during the course of the simulation. 
In model~7, as well as in most of the models considered in this study, 
we found that the evolution of the eccentricities of Jupiter and Saturn leads 
to values less than $0.1$. 
Therefore, the orbital configurations are generally stable during the crossing of 
(or trapping in) the MMR. These findings are in agreement with the stability study 
on the development of orbital eccentricity presented in \citet{ZhangZhou2010}.

\section{Masset \& Snellgrove Setup}
\label{sec:B}

\begin{figure}
 \includegraphics[width=0.8\linewidth]{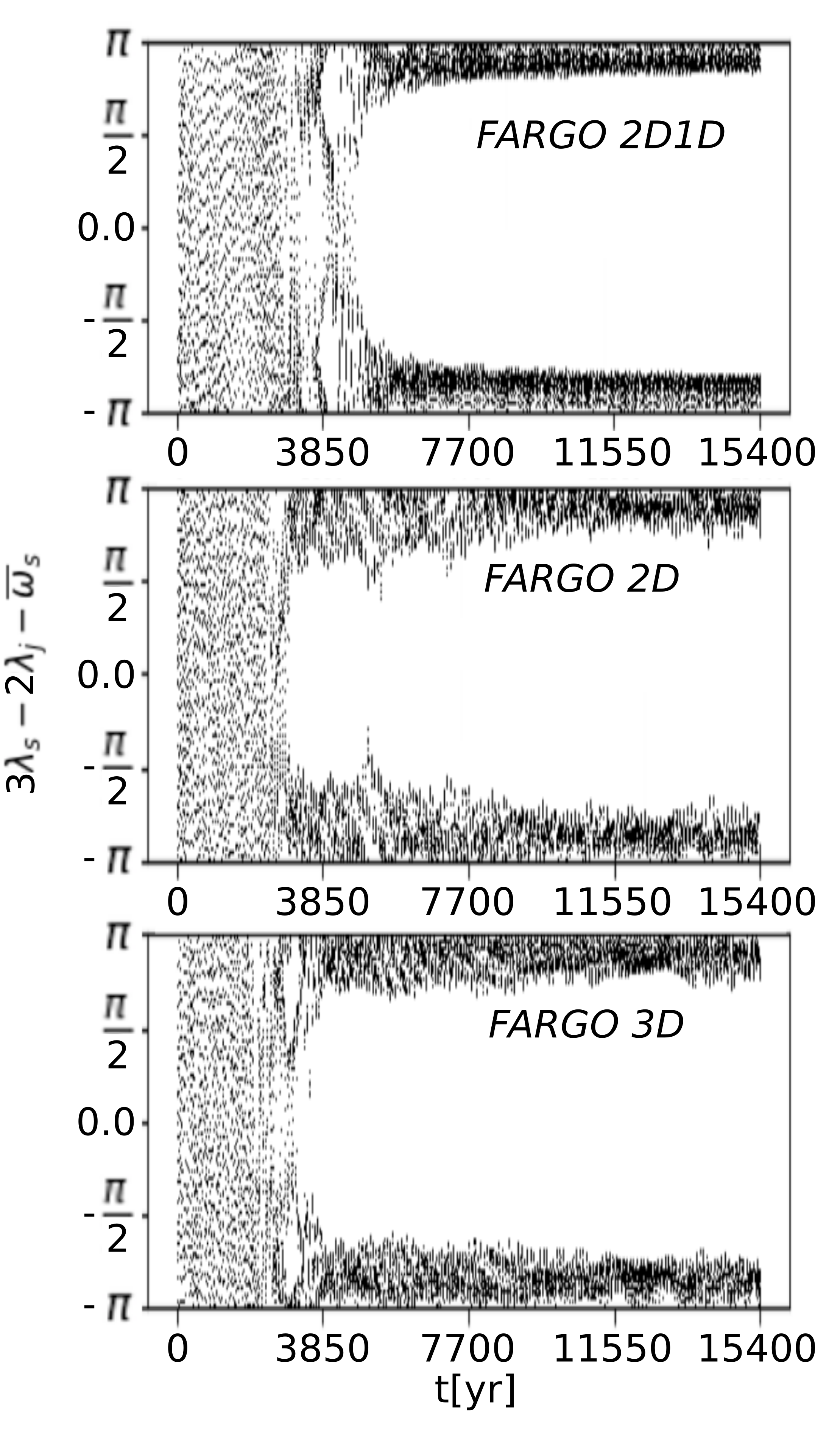}
    \caption{Resonant angles for the Masset $\&$ Snellgrove mechanism, obtained from different codes, as indicated.
             }
    \label{fig:raF}
\end{figure}
\begin{figure*}
 \includegraphics[width=\linewidth]{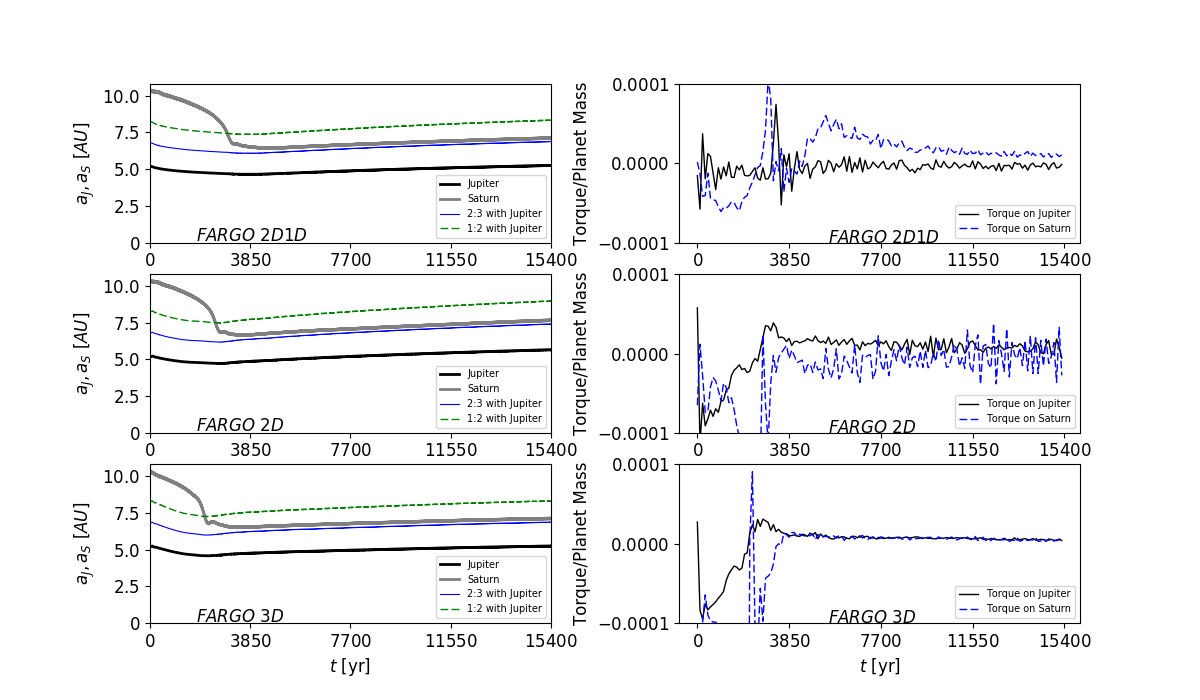}
    \caption{Left: Temporal evolution of the semi-major axis of the Jupiter and Saturn, trapped in the 3:2 MMR, obtained from calculations executed with three codes,
    as indicated. The setup replicates the parameters of \citet{MassetSnellgrove2001}. Right: Total torque on the Jupiter (black line) and Saturn (blue line) calculated from the density outputs of the calculationss in the left panels.
    }
    \label{fig:tqF}
\end{figure*}
Prior to carrying out the study presented above, we performed simulations to reproduce the outward migration results of \citet{MassetSnellgrove2001} using the FARGO2D1D code and considering the parameters given in section~2 of \citet{MorbidelliCrida2007}. In addition, we performed simulations of the Masset $\&$ Snellgrove mechanism with the codes FARGO2D and FARGO3D (results from these simulations are shown in Figures~\ref{fig:raF} and ~\ref{fig:tqF}). In order
to be sure that the planets are trapped in the 2:3 resonance we calculated the resonant angle $\varphi_R=3\lambda_s-2\lambda_j-\overline{w}_s$, where $\lambda_s$, $\lambda_j$ and $\overline{w}_s$ are the mean longitude of Saturn, the mean longitude of Jupiter and the longitude of Saturn's pericenter, respectively (see Figure~\ref{fig:raF}). We also computed the torques on the planets in these simulations.
We found that the total torque on Saturn and Jupiter calculated with FARGO2D1D is slightly different from that obtained from FARGO2D and FARGO3D. The difference may be due to two factors. 

First, the boundary conditions applied in the codes: while FARGO2D and FARGO3D apply a closed inner boundary condition, FARGO2D1D uses an open boundary condition between the 2D and the 1D meshes. The evolution of the disc inside Jupiter's orbits is therefore somewhat different, as FARGO2D1D is designed to better approximate the disc's viscous evolution at small radii. 

Second, the cut-off radius applied to the computation of the torque. In FARGO2D1D the torque in each cell of the 2D mesh is multiplied by the function
\begin{equation}
H_{cut}=1-\exp[-(d/R_{H})^2],
\end{equation}
where $d$ is the center-cell distance to the planet. 
Instead, in the FARGO2D and FARGO3D 
the torque is modulated by the 
cut-off function
\begin{eqnarray}
H_{cut} = \left \{ \begin{matrix} 0 & \mbox{if }\mbox{ $d/R_{H}<0.5$}
\\ 1 & \mbox{if }\mbox{ $d/R_{H}>1.0$} 
\\ \sin^2[\pi(\frac{d}{R_{H}}-\frac{1}{2})] & \mbox{otherwise.  }\mbox{}\end{matrix}\right. 
\end{eqnarray}

When modulating the torques in the simulations presented above, we chose to use the default cut-off function
implemented in FARGO2D1D \citep{MorbidelliCrida2007}. 
The difference in the values of the torque
is small overall and does not affect significantly the orbital evolution 
of the planets, as it can be observed in Figure~\ref{fig:tqF}.


\bsp	
\label{lastpage}
\end{document}